\documentclass[A4paper,english,aps,prb,twocolumn,groupedaddress]{revtex4-1}
\usepackage[LGR,T1]{fontenc}
\usepackage[latin9]{inputenc}
\usepackage{textcomp}
\usepackage{amssymb}
\usepackage{graphicx}
\usepackage{subscript}
\usepackage{amsmath}

\makeatletter

\DeclareFontEncoding{LGR}{}{}
\DeclareTextSymbol{\~}{LGR}{126}

\makeatother

\usepackage{babel}
\begin{document}

\title{In-plane angular dependence of superconducting gaps in FeSe probed by high resolution specific heat  measurements }

\author{H. Cercellier$^1$}
\email{herve.cercellier@neel.cnrs.fr}
\author{K. Olson$^1$}
\author{M. Schuchard$^1$}
\author{P. Toulemonde$^1$}
\author{A.-A. Haghighirad$^2$}
\author{C. Marcenat$^3$}
\author{T. Klein$^1$}

\affiliation{
$^1$ Univ. Grenoble Alpes, CNRS, Grenoble INP, Institut N\'eel, F-38000 Grenoble, France\\
$^2$ Institute for Quantum Materials and Technologies, Karlsruhe Institute of Technology, Kaiserstr. 12, 76131, Karlsruhe, Germany\\
$^3$ Univ. Grenoble Alpes, CEA, IRIG, PHELIQS, LATEQS, F-38000 Grenoble, France}

\date{\today}
\begin{abstract}
The influence of a rotating magnetic field (in the $ab-$plane) on the density of states has been investigated in the superconducting state of the nematic FeSe superconductor using high sensitivity specific heat measurements. As expected for (quasi-)nodal superconductors, oscillations in the specific heat ($C$) associated to the Doppler energy shift of Cooper pairs with momenta close to the gap minima are observed. In the $T_c = 9$~K crystal, $C(\phi$) displays a twofold symmetry at low temperature and low magnetic field confirming the nematic character of FeSe from thermodynamical measurements. As expected, a $\pi/2$ phase shift is observed for increasing temperatures (at $H=1$~T) but the gap structure abruptly changes above $\sim 1$~K in this sample. At low temperature, the maxima observed for $H||a-$axis at low $H$ split into lobes at $\pm 45^\circ$ when the magnetic field is increased indicating an anomalous field dependence of the gaps.

\end{abstract}

\maketitle

FeSe is a particularly interesting iron-based superconductor,  composed only of a c-axis stacking of FeSe layers without any charge reservoir. Despite its moderate critical temperature ($T_c \sim 9$~K) this compound can be seen as an "extremely" high $T_c$ material, lying at the verge of a Bose Einstein Condensation, due to its very low carrier density. The Fermi, superconducting gap and Zeeman energies (for fields on the order of the  $T=0$ upper critical field) are then on the same order of magnitude leading to an unprecedented superconducting state of highly spin-polarized electrons (see  [1-3]  for  extensive studies of the $H-T$ phase diagram). Moreover, in contrast to other iron-based materials, FeSe does not order magnetically \cite{Zhou} but superconductivity here competes with an orbitally ordered nematic state\cite{nematicity1,nematicity2,nematicity3} which breaks  the $C_4$ lattice symmetry down to $C_2$ below $\sim$ 90~K.  Angle-resolved photoemission spectroscopy studies evidenced strongly renormalized electron and hole dispersions, with a very anisotropic spectral weight in the normal state \cite{nematicity2,Watson1}. 

Unraveling the superconducting gap structure is then crucial to the understanding of this intriguing superconductor. Nematic order is expected to couple the $s$ and $d$ wave harmonics of the associated superconducting order parameter (of $s + d$ symmetry) but the amplitudes of the $s$ ($\Delta_s$) and $d$ ($\Delta_d$) components of the gaps are highly sensitive to nematicity and intra/interband coupling\cite{Islam}. As a consequence, accidental gap nodes (i.e. for angles which do not correspond to high-symmetry directions of the crystal) can appear if $\Delta_d>\Delta_s$.  It has finally been suggested that a transition from the $s + d$  to an $s+e^{i\alpha}d$ state could occur at low temperature for a (limited) range of s/d pairing interactions \cite{Islam,Kang}. However, even if it is now well established that the gaps present a twofold anisotropy with pronounced minima along $k_x$ and $k_y$ for the electrons and holes, respectively\cite{Cercellier,Kreisel,Sprau,Hardy,Rossler,Lin}, the presence of nodes remains controversial. Indeed, a nodal gap structure was inferred from both specific heat\cite{Hardy} (see discussion in [11]) and scanning tunneling spectroscopy \cite{Kasahara} measurements whereas nodeless superconductivity was rather supported by other specific heat \cite{Lin}  and thermal conductivity measurements \cite{Dong,Bourgeois}. Photoemission measurements \cite{Hashimoto} then showed that the gap of the hole pocket has a finite minimal value in multi-domain samples but gap values steeply dropping to zero in a narrow angle range in single domain samples.  

Specific heat ($C$) measurements in presence of a rotating in-plane magnetic field can then be an efficient mean in order to probe this gap structure into details. Indeed, in the mixed state, the flow of the Cooper pairs associated with the screening of the field in the vortex cores, locally shifts the energy required to create  unpaired quasiparticles (so-called Doppler shift). Rotating the magnetic field is then expected to lead to maxima (or minima) in the density of states\cite{Boyd}, when $\delta E_{\rm Dop} \propto \vec{v_s}.\vec{k_F}$ becomes on the order of the superconducting gap ($\vec{v_s}$ being the velocity of the screening currents). This (semiclassical) approach confirmed the oscillations previously inferred for intermediate magnetic fields within the so-called Brandt-Pesch-Tewordt approximation, taking into account the (spin-flip) scattering of the quasiparticles on vortices\cite{Vorontsov}, hence showing that $C/T(\phi)$  ($\phi$ being the angle of the magnetic field within the $ab$-pane) can provide fruitful information on the gap structure of nodal - or highly anisotropic - superconductors. 

 \begin{figure}
\begin{center}
\resizebox{0.45\textwidth}{!}{\includegraphics{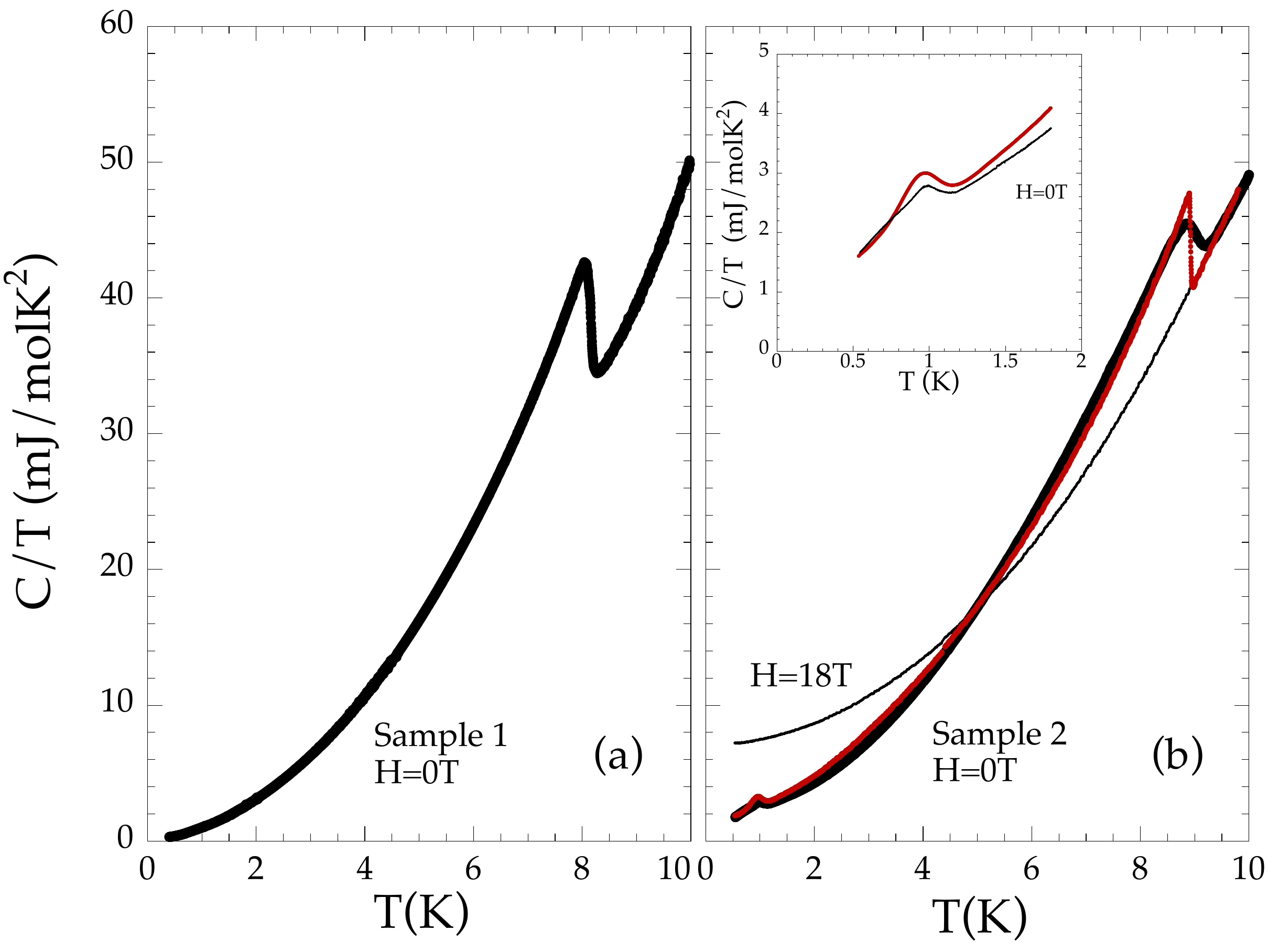}}
\caption{Temperature dependence of the specific heat divided by the temperature in a FeSe single crystal for $H=0$ (solid symbols) and $H=18$ T (thin line in Fig.1b). As shown, a second anomaly is clearly visible around 1~K in sample 2. This anomaly is well reproduced by our calculations introducing the drastic change in the gap structure inferred from the in-plane anisotropy data (see red line and zoom in inset). }
\end{center} 
\end{figure}

 \begin{figure}
\begin{center}
\resizebox{0.48\textwidth}{!}{\includegraphics{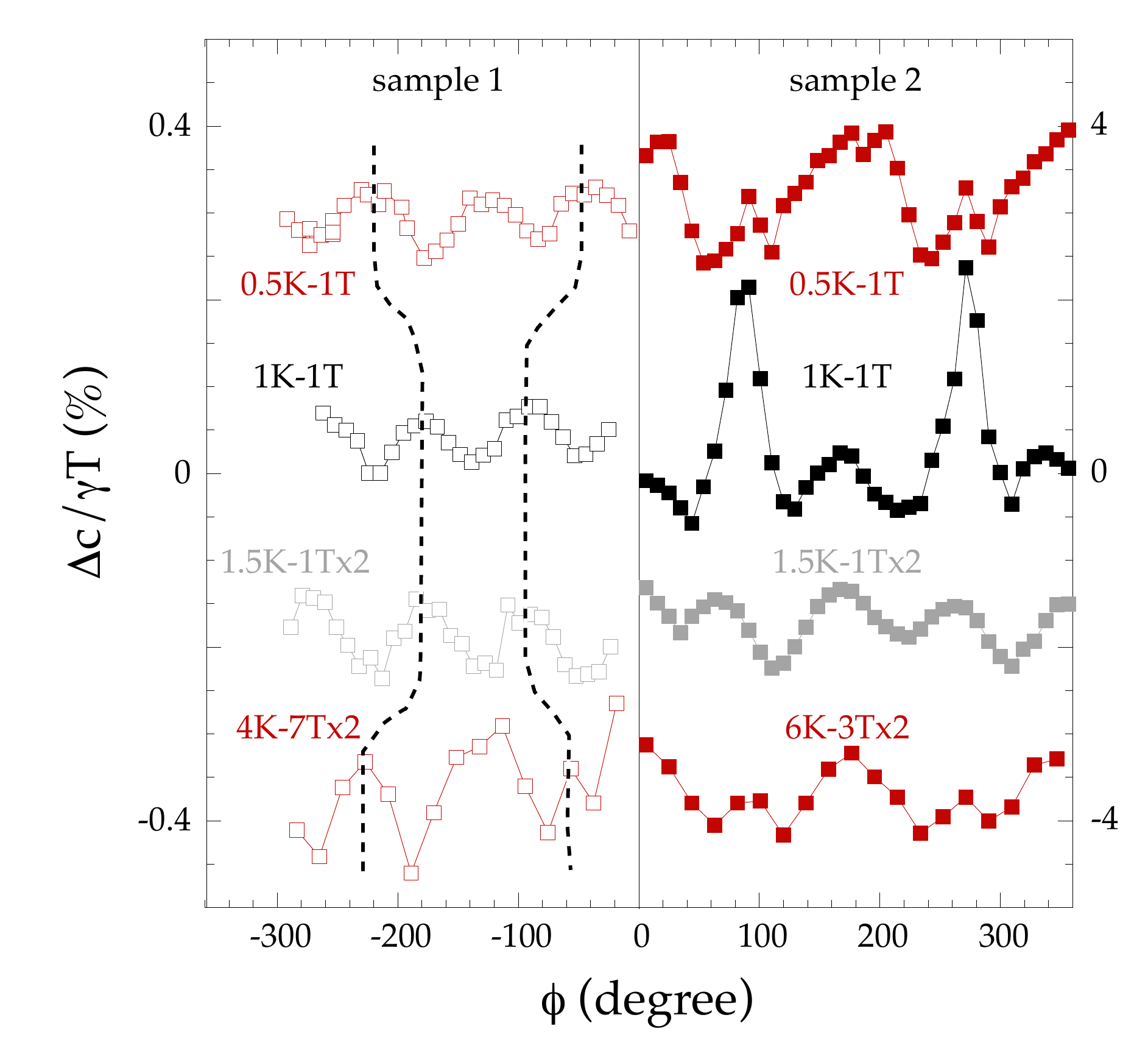}}
\caption{Angular oscillations of the specific heat at $H=1$~T for rotating magnetic fields within the $ab-$plane in sample 1 (open symbols, left column) and sample 2 (closed symbols, right column) at the indicated temperatures (the different curves have been arbitrarily shifted for clarity). The $\phi=0^\circ$ angle corresponds to the field aligned with the crystal edge i.e. the [100] direction of the low $T$ orthorhombic phase (as deduced from X-ray diffraction measurements on sample 1). As shown a $C_2$ symmetry is clearly visible in sample 2 confirming the nematic character of FeSe (note the dramatic fall of the modulation amplitude above 1~K, see also Fig.3). The amplitude of the modulations is strongly reduced in sample 1 (of lower $T_c$) which only displays a $C_4$ symmetry. Dotted lines are guides to the eyes underlying the 45$^\circ$ phase shifts of the oscillations (see text for details). As shown oscillations could be observed up to high temperatures (and/or magnetic fields, see also Fig.3) and two {\it inversions} of the oscillations are visible (red to black colors and dotted guides to the eyes).}
\end{center} 
\end{figure}

 \begin{figure*}
\begin{center}
\resizebox{1\textwidth}{!}{\includegraphics{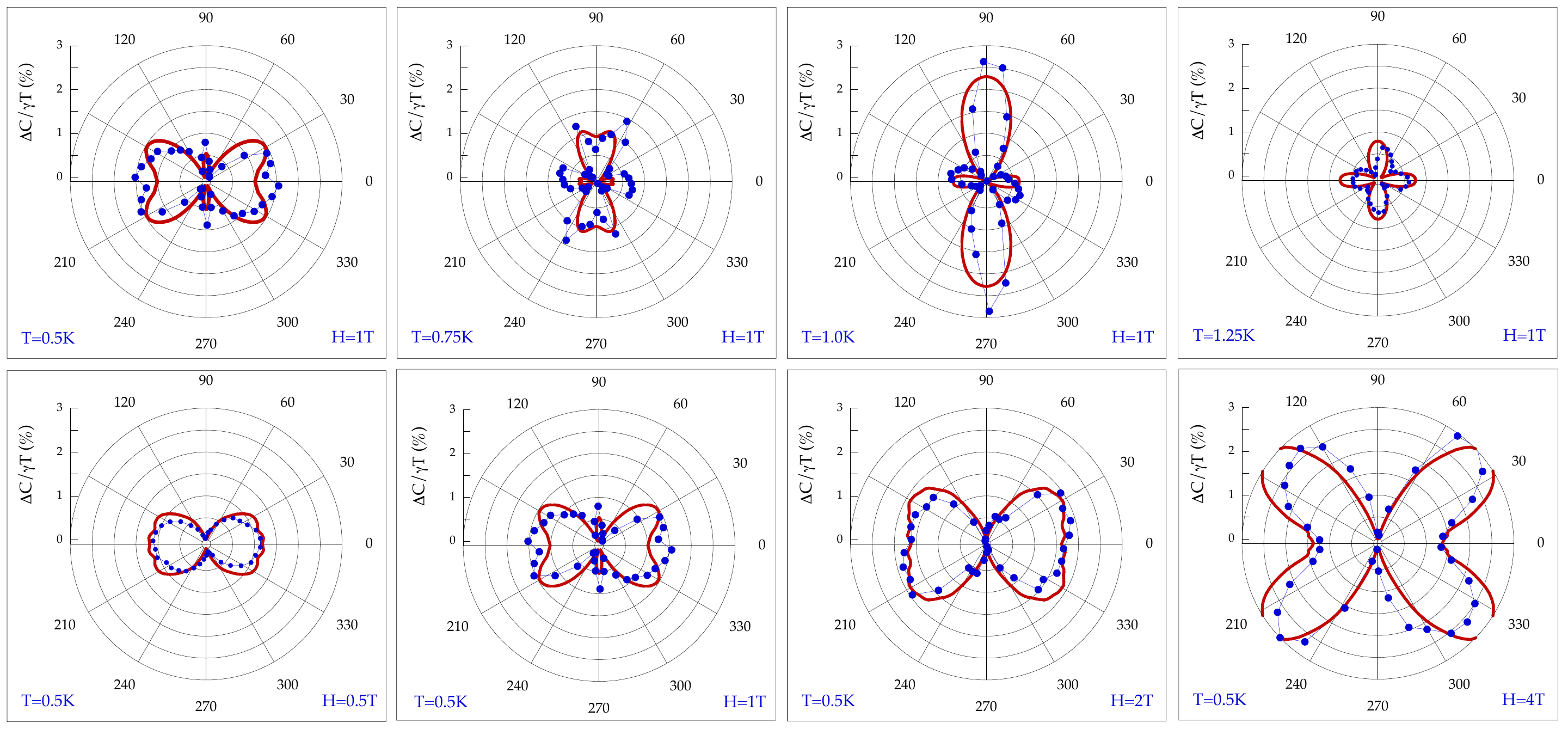}}
\caption{Polar plot of the angular dependence of the specific heat (after subtraction of its minimal value) for rotating magnetic fields within the $ab-$plane in FeSe (sample 2) at the indicated fields and temperatures (first row : temperature dependence at $H=1$~T, second row : field dependence at $T=0.5$~K). The solid red lines are numerical calculations of the $C(\phi)/T$ modulations in presence of the Doppler shift using the gap structure displayed in Fig.4 and the electronic structure given in [19] (see [24] for additional data).   }
\end{center} 
\end{figure*}

Modulations in $C/T(\phi)$ have been reported by Sun {\it et al.} \cite{Sun1} in a $T_c\sim 9$~K FeSe crystals, confirming the strong anisotropy of the gaps. However, the amplitude of the $C/T$ modulations are very small ($\approx \gamma/100$ at best, where $\gamma$ is the Sommerfeld coefficient) and they could only be observed in a limited $T$ and $H$ range ($H \lesssim 2$~T and $T\lesssim 2$~K). Moreover,  those preliminary measurements could not reveal the nematic character of FeSe due to strong twinning of the sample, hindering any quantitative  analysis of the gap structure. The size of the domains is ranging from (typically) $\sim$10 $\mu$m to $\sim$100 $\mu$m (in the absence of external strain, see for instance photoemission data \cite{Hashimoto}). in the $C/T$ measurements, the samples are slightly strained at low $T$ by the solidification of the grease (slightly) favoring single domains but  miniature samples are still required. Two ($\sim 10^{-2}$ mm$^3$) single crystals with different $T_c$ values (8.1~K and 9.0~K in sample 1 and 2, respectively) were hence selected to study the angular dependence of $C/T$ using a very high resolution specific heat set-up  (see [24] and [2] for further details).  

Oscillations could be resolved up to 6~K and/or 7~T in both samples (see Fig.2 and 3 below) and a clear twofold symmetry in the $C/T$ in plane anisotropy was observed at low temperature and low magnetic field in sample 2 (see Fig.3 below for $T=0.5$~K and $H = 0.5$T), clearly indicating that this sample is only weakly twinned and directly confirming the nematic character of FeSe from thermodynamical measurements. Moreover, our measurements reveal a complex angular dependences of the specific heat for higher $T$ and/or $H$ values (see Fig.2 and 3). This anisotropy can  be very well reproduced by numerical calculations taking into account the Doppler shift effect (see discussion), revealing a complex $T$ and $H$ dependence of the gap structure. Note that  we  verified that no oscillations in $C/T$ are present in the empty chip down to 0.5~K and up to 7~T, and checked  for the good alignment of the samples within the ab-plane \cite{SM}. At 6~K, $\Delta C/C_{\rm tot} \sim 5.10^{-4}$ ($C_{tot}$  being the total specific heat including phonons and addenda) and the data for each angle have been averaged over approximatively 10 minutes to reach the required sensitivity ($\sim 5.10^{-5}$). 

 As shown in Fig.1, both samples (synthesized by chemical vapor transport, see [11,15] and [25,26] for further information on the synthesis and first characterisation) present well resolved specific heat anomalies at $T_c$ attesting for their good quality (the transition is even significantly sharper in sample 1 despite its lower $T_c$ value). A second anomaly is also clearly visible around $T\sim 1$K in sample 2 (see Fig.1b). A bump in the temperature dependence of $C/T$ has been  previously reported in several studies \cite{Sun2,Hardy,Rossler} and has initially been attributed to the gap anisotropy in the $s+d$ wave model. However, such anisotropy cannot account for the existence of a {\it jump} in $C/T$ \cite{Chen}. It was then suggested that this jump might be attributed to the $s+e^{i\alpha}d$ transition mentioned above\cite{Kang} but further theoretical works led to the conclusion that this scenario is unlikely as this transition only shows up in a narrow parameter range\cite{Islam}. We will show here that this anomaly is inferred by a rapid change of the gap structure around 1~K. It is worth noting that it was observed in most of the samples of batch 2 (but not all) and could even reach $\sim 4$ mJ/molK$^2$ in some (rare) cases\cite{SM}, clearly indicating that its presence/amplitude is very sensitive to minor changes from sample to sample. 

 \begin{figure}
\begin{center}
\resizebox{0.48\textwidth}{!}{\includegraphics{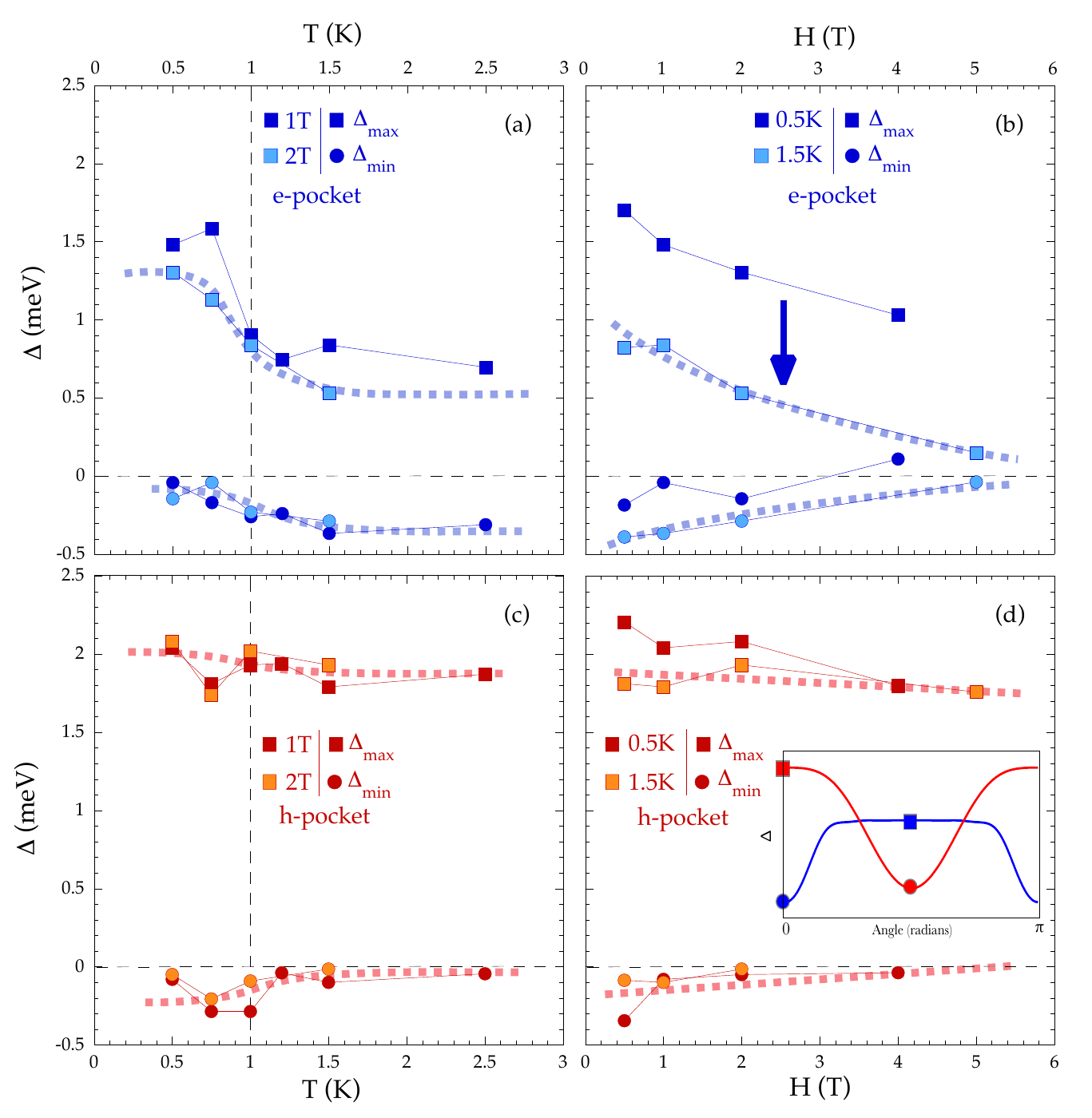}}
\caption{Temperature (at $H=1$~T (closed symbols) and 2~T (open symbols), panel a and c) and field (at $T=0.5$~K (closed symbols) and 1.5~K (open symbols, panel b and d) dependence of the gap maxima (squares) and minima (circles) for the hole (red symbols) and electron (blue symbols) pockets of the Fermi surface, used to fit the angular dependence of the specific heat (solid red lines in Fig.3). The inset of Fig.4d displays the gap functions. As shown, a drastic change in the gap structure is observed for $T \sim 1.1$~K (see dashed guide to the eye at 2~T) for the electron pocket, where as the hole pocket remains almost temperature independent. For this electron pocket both $\Delta_{\rm max}$ and $\Delta_{\rm min}$ rapidly decrease with field (see dashed guide to the eye at 1.5~T) giving rise to lobes in the angular dependence of $C/T$ (see Fig.3). Note that this gap seems to vanish above  $\sim$ 1.5~K / 5~T. }
\end{center} 
\end{figure}

The temperature and field dependence of $C/T(\phi)$ in sample 2 is displayed in Fig.3 (see also additional data in supplemental materials\cite{SM}).  As shown a "simple" $C_2$ symmetry is observed at low $T$ and $H$ but much more complex patterns show up for increasing $T$ and/or $H$. For every set of temperature and magnetic field, we performed numerical calculations of the $C/T(\phi)$ modulations in presence of the Doppler shift. The calculations are based on the model band structure used in \cite{Cercellier}, taking into account the orbital-dependent spectral weight and introducing the wave-vector dependent Doppler shift following the formalism of Graser \cite{Graser} valid for small temperatures and magnetic fields (see supplemental materials \cite{SM} for details). The experimental $\Delta C/\gamma T$ curves have  been fitted to the calculated  ones after subtraction of the standard phonon and electronic contributions. As shown (see solid red lines in Fig.3) a very reasonable quantitative agreement between calculations and experimental data has been obtained for all $T$ and $H$ values, by adjusting the hole and electron gap extrema (see Fig. 4). The minor discrepancies are much probably due to slight twinning, slight misalignment of the sample and/or "details" of the electronic structure.

As pointed out by Boyd {\it et al.} \cite{Boyd}, the angular dependence of the specific heat oscillations at finite $T$ and $H$ does not merely reflect the low-energy density of states of the superconductor, but involves the whole density of states up to the coherence peak energy. This can lead to an inversion of the maxima and minima of the specific heat depending on temperature and field, here well reproducing the $90^\circ$ rotation of the $C/T$ maxima observed around 1~K.  Note that accidental nodes can appear in the $s+d$ symmetry for cos$(2\Psi)=-\Delta_s/\Delta_d$ ($\Psi$ being the angle along the Fermi pocket \cite{Chubukov}) and the location of the maxima and minima can hence not be directly anticipated  in this system. Calculations based on the full band structure and gap functions are necessary as $C/T(\phi)$ then clearly differs from a simple cos($4\phi$) function. Those calculations well reproduce the intriguing "butterfly" structure observed at $T=0.75$K (and $H= 1$~T), clearly indicating the presence of accidental nodes in the gap structure, for both the electron and hole pockets (in good agreement with the angle resolved photoemission measurements \cite{Hashimoto}).

The as-deduced gap values have been reported in Fig.4. Those values are in very reasonable agreement with those previously obtained from spectroscopic measurements \cite{Sprau} (see also discussion in [14]) but our measurements unveil an unusual $T$ and $H$ dependence of the gaps. Indeed, the gap values only slightly depend on $T$ above $\sim 1$~K  (see Fig.4a) but changes strikingly below 1~K for the electron pocket : the maximum abruptly increases from $\Delta_{\rm max}^{\rm electron}\sim 0.7$ meV  to  $\sim 1.5$ meV  and $\Delta_{\rm min}^{\rm electron} \rightarrow 0$ meV (the hole gap structure only weakly changes\cite{SM} with mainly a slight decrease of $\Delta_{\rm min}^{\rm hole}$). It is then worth noting that this drastic change in the gap structure accounts for the change observed in the in-plane specific heat anisotropy but also directly accounts for the anomaly at 1K. Indeed, the $d\Delta^2/dT$ contribution to $C/T$ gives rise to  a  "peak" in the temperature dependence of $C/T$ which consistently reproduces the 1K-anomaly observed in sample 2 (see red line in Fig.1b (and zoom on 0-2~K range in the inset) and Fig.2-SM in [24]), hence directly providing an explanation for this anomaly. 

Similarly, whereas the gap structure of the hole pocket only slightly depends on $H$ (see Fig.4d and [24] for additional data), both gap maxima and minima rapidly decreases with field for the electron pocket (see Fig4b). As a consequence, lobes develop in $C/T(\phi$) at $\pm 45^\circ$ at low temperatures leading to an effective fourfold symmetry at 0.5~K/4~T (see Fig.3).  It is  also worth noting that a rapid increase of the specific heat has been reported at low temperature below 2-3~T \cite{Sun1,Hardy}. This increase is very similar to the one previously observed in MgB$_2$ due to the presence of both a robust $\sigma-$gap (closing at $H_{c2}$) and a fast closing $\pi$-band gap, vanishing for magnetic fields $\sim H_{c2}/10$ \cite{MgB2}. This rapid increase of $C/T$ in FeSe is then consistent with the closing of the electron gap (vanishing above $\sim 1.5$~K/5~T) in presence of a more robust hole gap. However, to the best of our knowledge, there is no straightforward interpretation for such a field dependence of the gaps  in FeSe. Note that it has also been shown that the mixed state of FeSe exhibits a vortex-lattice transformation from a nearly hexagonal to a nearly square lattice for magnetic fields between 1T and 4T (applied along the $c$-axis) which could result from the interplay between nematicity and the $s+d$ order parameter \cite{Lu,Putilov} but the  influence of this interplay on the field dependence of the gaps still has to be clarified. 

Let us finally discuss the influence of $T_c$. Earlier works\cite{Cercellier,Sun2} pointed out a significant decrease of the gap maxima with $T_c$ and suggested that nodes could be wiped out in lower $T_c$ samples. The origin of this change of the gap structure with $T_c$ remains however unclear. In contrast to earlier measurements\cite{Sun2}, we show that $C/T$ modulations are also present in our $T_c = 8.1$~K sample (sample 1, left panel of Fig.2), but reduced in amplitude. The anisotropy measured above 1~K is qualitatively similar to the one observed in sample 2 (see Fig.2) but, in contrast to what observed in sample 2, the modulations remain small down to the lowest temperatures (being then reduced by roughly an order of magnitude as compared to sample 2). This suggests that the gap structure only slightly changes with $T$ in this sample. Concomitantly, no "1K-anomaly" was observed in any of the samples of batch 1. Unfortunately, a $C_4$ symmetry was observed for all $T$ and $H$ in this sample, suggesting the presence of larger twinning hence hindering any quantitative analysis. However, it is worth noting that maxima at 45$^\circ$ are observed in sample 1  at 1~T (see dotted line in Fig.2). As discussed above, the lobes observed at 45$^\circ$ in sample 2 (at 4~T /0.5~K) result from the decrease of the electron gap with field and, similarly, the 45$^\circ$ $C/T$ maxima observed in sample 1 (at lower fields) could hence be due to the decrease of the electron gap with $T_c$  in this sample. 

Finally, note that the Doppler shift approximation used here is only valid when vortices are far apart ($H<<H_{c2}$) and for energies small compared to the maximal superconducting gap hindering calculations at high temperatures or magnetic fields. Still it is worth noting that a $C/T$ modulation very similar to the one observed at low~$T$ / low$H$  is recovered  at high~$T$ / high~$H$ (see Fig.2 and [24]). We already mentioned the presence of a $90^\circ$ rotation of the $C/T$ maxima at low temperature, this {\it second inversion} - also observed in sample 1 - could then be reminiscent of the double phase shift obtained by Vorontsov {\it et al.} \cite{Vorontsov}  in the Brandt-Pesch-Tewordt approximation (for $T/T_c \sim 0.1$ and $0.5$, in reasonable agreement with the present data).

In summary, we have shown that the angular dependence of $C/T$ of FeSe crystals for rotating magnetic fields within the ab-plane can be reproduced very well by numerical calculations taking into account the Doppler shift. For samples of optimal $T_c$ values ($\sim 9$~K), our data confirm the nematic properties of FeSe from thermodynamic measurements and show that the gap structure clearly exhibits accidental nodes associated to the $s+d$ symmetry of the order parameter. While the gap of the hole pocket only weakly depends on $T$ (and $H$), a rapid change in the gap structure is observed for $T \sim 1$K for the electron pocket giving rise to a jump in the temperature dependence of $C/T$. Moreover, the amplitude of both the gap maxima and minima of this pocket rapidly decrease with field, leading to an effective fourfold symmetry of the $C/T$ modulations above $\sim 4$~T at low temperature. The amplitude of the $C/T(\phi)$  modulations is decreased by about an order of magnitude in samples of lower $T_c$ values, indicating that the gap structure is highly sensitive to $T_c$. The physical origin of this anomalous $T$ and $H$ dependences of the gap structure remains to be explained but this interpretation falls beyond the scope of the present work.

Acknowledgments. The authors thanks J\'er\^ome Debray and Jacques P\'ecaut for their Laue X-ray diffraction and 4-circle XRD measurements, respectively.
 
\bibliographystyle{apsrev4-1}

 \end{document}